\documentstyle[epsfig]{article}
\def\bib{\parskip=0pt\par\noindent\hangindent\parindent
    \parskip =2ex plus .5ex minus .1ex}

\newcommand{\be}{\begin{equation}}
\newcommand{\ee}{\end{equation}}
\newcommand{\ba}{\begin{eqnarray}}
\newcommand{\ea}{\end{eqnarray}}

\newcommand{\nn}{\nonumber \\}
\newcommand{\nnb}{\begin{displaymath}}
\newcommand{\nne}{\end{displaymath}}
\newcommand{\de}{\partial}

\newcommand{\lb}{\mbox{\boldmath $\ell$}}
\newcommand{\lbh}{\mbox{\boldmath $\hat{\ell}$}}

\newcommand{\r}{\mbox{\boldmath $r$}}

\newcommand{\thetab}{\mbox{\boldmath $\theta$}}

\newcommand{\hth}{\mbox{\boldmath $\theta$}}
\newcommand{\om}{\Omega_m}
\newcommand{\fsky}{f_{\rm sky}}
\newcommand{\ol}{\Omega_v}
\newcommand{\0}{\mbox{\boldmath $0$}}

\newcommand{\rhat}{\hat{\r}}

\newcommand{\Mpc}{\, h^{-1}{\rm Mpc}}

\newcommand{\rgl}{\rangle}
\newcommand{\lgl}{\langle}

\begin{document}
\centerline{\huge\bf New Dimensions in Cosmic Lensing}

\vspace{0.5cm} \centerline{\Large\bf Andy Taylor} \vspace{0.15cm}
\centerline{	Institute for Astronomy, 
		School of Physics,
 		University of Edinburgh,}
\centerline{  	Royal Observatory Edinburgh, 
		Blackford Hill, }
\centerline{	Edinburgh, EH9 3HJ, U.K.}
\centerline{    email: ant@roe.ac.uk}
\hspace{1.in}

{\bf I review the current status of combing weak gravitational lensing 
 with depth information from redshifts
 as a direct probe of dark matter and dark energy in the Universe.
 In particular I highlight:
 (1) The first maximum likelihood measurement of the cosmic shear power
 spectrum, with the COMBO17 dataset (Brown et al 2003);
 (2) A new method for mapping the 3-D dark matter distribution  
 from weak shear, and its first application to the COMBO17 dataset
 (Taylor et al 2003);
 (3) A new method for measuring the Dark Energy of the Universe
 using purely the geometry of gravitational lensing, based on 
 cross-correlation tomography (Jain \& Taylor 2003). I show that this 
 method can constrain the
 equation of state of the universe and its evolution to a few percent
 accuracy.
}


\section{Introduction}

Gravitational lensing provides us with the most direct and
cleanest of methods for probing the distribution of matter in the
universe (Mellier 1999, Bartelmann \& Schneider 2001). The lensing
effect arises due to the scattering of light by perturbations in
the metric, stretching and contracting bundles of light rays,
causing the distortion of background galaxy images. Hence
gravitational lensing does not depend on any assumptions about the
state of the intervening matter. These distortions manifest
themselves as a shear distortion of the source galaxy image (see
e.g. Tyson et al  1990; Kaiser \& Squires 1993), or a change in
the surface number density of source galaxies due to magnification
(see e.g. Broadhurst, Taylor \& Peacock 1995; Fort, Mellier \&
Dantel-Fort 1997; Taylor et al 1998) and can be used to map the
two dimensional projected matter distribution of cosmological
structure. As the matter content of the universe is dominated by
non-baryonic and non-luminous matter, gravitational lensing is the
most accurate method for probing the distribution of this dark
matter.

Weak lensing studies have been carried out for a wide range of 
galaxy clusters, allowing precision measurements of cluster masses
and mass distributions (see e.g. Tyson et al 1990, Kaiser \& Squires
1993, Bonnet et al 1994, Squires et al 1996, Hoekstra et al 1998,
Luppino \& Kaiser 1997, Gray et al 2002). On larger scales, the
shear due to large-scale structure has been accurately measured by
several groups (see e.g. van Waerbeke et al 2001, Hoekstra et al 2002, Bacon
et al 2002, Jarvis et al 2003, Brown et al 2003).

Depth information, from galaxy redshifts, has already been used in 
weak lensing studies to determine the median redshifts of the lens
and background populations. This can be a limiting factor in the analysis,
leaving uncertainty in the overall mass normalisation.  
In this review I describe an application using accurate
shear and redshift information from the COMBO17 dataset (Wolf et al, 2003)
to estimate the first maximum likelihood shear power spectrum analysis 
(Section 2), including all of the uncertainties associated with source
distances (see Brown et al 2003).
	While the shear power contains much important information on 
 the statistics of the dark matter distribution and cosmological
 parameters, a lot of information is projected out. However this is 
 not a necessary step. In Section 3 I outline how the full 3-D dark
 matter distribution can be recovered from shear and redshift 
 information, and make the first application to the COMBO17 data
 (see Taylor 2001, Bacon \& Taylor 2003, Taylor et al 2003).
	Finally, in Section 4, I describe a new method for using 
 shear and redshifts in a purely geometrical test 
 to accurately measure the equation of state of the 
 dark energy in the universe and its evolution (See Jain \& Taylor 2003).

	It is clear that the combination of shear and redshift 
 information provides a powerful tool for cosmology, allowing us
 to not only image the dark matter directly, but also see its evolution
 over cosmic time. Hence future lensing surveys should consider 
 being coupled to photometric redshift surveys, opening up 
 new dimensions in gravitational lensing studies.

\subsection{Basic weak lensing equations}

 The metric of a perturbed
 Robertson-Walker universe in the conformal Newtonian gauge is
 \be
    ds^2 = - (1+2 \Phi)dt^2 + a^2(t) (1-2 \Phi) dr_i dr^i,
 \ee
 where $\Phi$ is the Newtonian potential, $a$ is the cosmological
 scale factor, and we have assumed a
 spatially flat universe. The Newtonian potential
 is related to the matter density field by Poisson's
 equation
  \be
    \nabla^2 \Phi = 4 \pi G \rho_m \delta a^2 = \frac{3}{2}
    \lambda_{H}^{-2} \Omega_m a^{-1} \delta,
  \ee
  where $\delta = \delta \rho_m/\rho_m$ is the matter density
  perturbation, $\lambda_H=1/H_0 \approx 3000 \Mpc$ is the Hubble
  length, and $\Omega_m$ is the present-day mass-density parameter.
 The lensing potential, $\phi$, for a source in a
 spatially flat universe at a radial distance $r$ is given by
 (e.g.  Bartelmann \& Schneider 2001)
 \be
    \phi(r,r \thetab) =  2 \int_0^{r}\!\! dr' \,
     \left(\frac{r-r'}{r r'}\right) \Phi(r',r'\thetab),
 \label{pot}
 \ee
 at an angular position $\thetab$ on the sky. Hence the
 lensing potential is just a weighted radial projection of the 3-D
  gravitational potential.

In lensing the symmetric, tracefree shear matrix,
$\gamma_{ij}$, which describes the distortion of the lensed image,
and the magnification, $\mu$, which describes the change in area,
are observables. The shear matrix is
 \be
    \gamma_{ij} = \left(\de_i \de_j - \frac{1}{2} \delta^K_{ij}
    \de^2\right)\phi,
 \ee
 where $\de_i \equiv r (\delta_{ij}- \theta_i \theta_j) \nabla_j =
r(\nabla_i - \theta_i \de_r)$ is a dimensionless, transverse
differential operator, and $\de^2 \equiv \de_i \de^i$ is the
transverse Laplacian. The indices $(i,j)$ take the values $(1,2)$,
assuming a flat sky.
The lens convergence, $\kappa$, is defined by
 \be
    \kappa = \frac{1}{2} \de^2 \phi.
 \label{kap}
 \ee
 An estimate of the lensing potential, $\widehat{\phi}$, can be
 found from the shear field by the Kaiser-Squires (1993) relation;
 \be
        \widehat{\phi}(r,r\thetab) = 2  \de^{-4} \de_i \de_j \,
        \gamma_{ij} (r,r\thetab),
 \label{kaiser-squires}
 \ee
 where $\de^{-2}\equiv \int \!d^2\theta'\, \ln|\thetab-\thetab'|/2\pi$
  is the inverse 2-D Laplacian operator.
 In practice the shear field is only discretely sampled by
 galaxies so we must smooth the shear field to perform the
 differentiation. This also serves to make the uncertainty on the
 measured shear field finite, since each source galaxy has an unknown
 intrinsic ellipticity. The scalar convergence field, $\kappa$, can be 
 estimated from the shear field 
by combining equations (\ref{kaiser-squires}) and (\ref{kap}),
 up to a constant of integration.

 In principle there is another odd-party pseudoscalar which can be formed
 from a shear field;
 \be
    \beta = \varepsilon^m_{i} \de_{j} \de_m  \de^{-2} \gamma_{ij},
 \ee
 where $\varepsilon^i_j$ is the 2-D antisymmetric Levi-Civita tensor.
 This field cannot correspond to the even-parity lens convergence field
 and can be used to investigate noise, boundary effects and intrinsic
 alignments of galaxies (Catelan, Kamionkowski \& Blandford, 2001; Crittenden,
 et al 2001; Croft \& Metzler, 2001; Heavens, Refregier \& Heymans, 2000),
 which will appear in both
 $\kappa$ and $\beta$ modes. However, alignments appear to be a small effect
 at large redshift (Brown et al 2002, Heymans et al 2003).

\section{The Shear Power Spectrum}

\subsection{Statistical properties}

We may define a shear covariance matrix by
 \be
    C_{ab}(\thetab) = \lgl \gamma_a(\0) \gamma_b(\thetab) \rgl,
\label{shear_signal}
 \ee
where the indices $(a,b)$ each take the values (1,2). 
Fourier transforming the shear field,
 \be
    \gamma_{ij} (\lb) = \int \! d^2 \theta \,
        \gamma_{ij}(\thetab) e^{- i \lb . \thetab },
 \ee
and decomposing it into  $\kappa$ and $\beta$
we may generate the shear power spectra:
 \ba
    \lgl \kappa(\lb) \kappa^*(\lb') \rgl &=& (2 \pi)^2 C^{\kappa
    \kappa}_\ell \delta_D(\lb - \lb'), \nn
       \lgl \beta(\lb) \beta^*(\lb') \rgl &=& (2 \pi)^2 C^{\beta
    \beta}_\ell \delta_D(\lb - \lb'), \nn
       \lgl \kappa(\lb) \beta^*(\lb') \rgl &=& (2 \pi)^2 C^{\kappa
    \beta}_\ell \delta_D(\lb - \lb').
 \ea
The parity invariance of weak lensing suggests that
$C^{\beta\beta}_\ell=C^{\kappa\beta}_\ell=0$. However other
effects, such as noise and systematics, as well as intrinsic
galaxy alignments may give rise to a non-zero
$C^{\beta\beta}_\ell$.  The cross-correlation of 
$\kappa(\ell)$ and $\beta(\ell)$ is expected to be zero but it 
allows a second check on noise and systematics in the
shear field. In
particular finite field and boundary effects can lead to leakage of power
between these three spectra.
The shear power spectrum and the convergence power are related by
  $
         C^{\gamma \gamma}_\ell = C^{\kappa\kappa}_\ell
  $
in the flat-sky approximation. For a spatially flat Universe, 
these are in turn related to the matter power spectrum, 
$P_{\delta}(k,r)$ by the integral relation (see e.g. Bartelmann 
\& Schneider 2001):
  \be
	C^{\kappa \kappa}_\ell = \frac{9}{4} \left(\frac{H_0}{c}\right)^4 
	\Omega_m^2\, \int_0^{r_{\mathrm H}}\! dr \,
	P_\delta \left (\frac{\ell}{r},r\right)
	\left(\frac{\overline{W}(r)}{a(r)}\right)^2 ,
	\label{eq:gampowspec}
  \ee
where $a$ is the expansion factor and $r$ is comoving distance.
$r_{\rm H}$ is the comoving distance to the horizon:
  $
	r_{\rm H}= c \int_0^\infty dz/H(z),
  $  	
where the Hubble parameter is given in terms of the matter density, 
$\Omega_m$, the vacuum energy density, $\Omega_V$ and the spatial
curvature, $\Omega_K$ as
  $
	H(z)=H_0[(1+z)^3 \Omega_m+(1+z)^2\Omega_K+\Omega_V]^{1/2}.
  $
The weighting, $\overline{W}$, is given in terms of the normalised
source distribution, $G(r) dr = p(z) dz$:
 \be
   \overline{W}(r) \equiv \int_r^{r_{\mathrm H}}\! dr'\,G(r')\,
   \frac{r'-r}{r'}.
 \ee
The covariances of the shear 
field, equation (\ref{shear_signal}), are related to
the power spectra by (Hu \& White 2001)
 \ba
    C_{11}(\rhat) &=& \int \! \frac{d^2 \ell}{(2 \pi)^2} \,
    \big( C^{\kappa\kappa}_\ell \cos^2 2 \varphi_\ell +
            C^{\beta \beta}_\ell \sin^2 2 \varphi_\ell  -
            C^{\kappa \beta}_\ell \sin 4 \varphi_\ell \big)|W(\ell)|^2
            e^{i \lb.\rhat}, \nn
    C_{22}(\rhat) &=& \int \! \frac{d^2 \ell}{(2 \pi)^2} \,
    \big( C^{\kappa\kappa}_\ell \sin^2 2 \varphi_\ell +
            C^{\beta \beta}_\ell \cos^2 2 \varphi_\ell +
            C^{\kappa \beta}_\ell \sin 4 \varphi_\ell \big)|W(\ell)|^2
            e^{i \lb.\rhat}, \nn
    C_{12}(\rhat) &=& \int \! \frac{d^2 \ell}{(2 \pi)^2} \,
    \big( \frac{1}{2}(C^{\kappa\kappa}_\ell -
            C^{\beta \beta}_\ell) \sin 4 \varphi_\ell +
            C^{\kappa \beta}_\ell \cos 4 \varphi_\ell \big)|W(\ell)|^2
            e^{i \lb.\rhat},
 \label{shearcovar}
 \ea
 where $\cos \varphi_\ell = \lbh . \lbh_x$ and $\lbh_x$ is a
 fiducial wavenumber projected along the x-axis. We have included here
 a smoothing or pixelisation window
 function;
 $
    W(\lb) = j_0(\ell_x \theta_{\rm pix}/2)j_0(\ell_y \theta_{\rm pix}/2),
 $
 where $j_0=\sin(x)/x$ is the zeroth order spherical Bessel
 function and $\theta_{\rm pix}$ is the smoothing/pixel scale.

 With these relations, the three shear power spectra $C^{\kappa\kappa}_\ell$, 
$C^{\beta \beta}_\ell$ and $C^{\kappa \beta}_\ell$ can be estimated 
directly from shear data via a Maximum Likelihood approach (Hu \& White 2001,
Brown et al 2003). The dataset we have applied it to is the COMBO17 survey.

\subsection{The COMBO17 Dataset}

The data analysed is part of the COMBO-17 survey
(Wolf et al. 2001), carried out with the Wide-Field Imager (WFI) at the MPG/ESO
2.2m telescope on La Silla, Chile. The survey currently consists of five 
$0.5^\circ \times 0.5^\circ$ 
fields totaling 1.25 square degrees with observations taken in five 
broad-band filters ($UBVRI$) and 12
narrow-band filters ranging from $420$ to $914$ nm. The chosen filter 
set facilitates accurate
photometric redshift estimation ($\sigma_z \approx 0.05$) reliable 
down to an $R$-band magnitude of 24.
During the observing runs the best seeing conditions
were reserved for obtaining deep $R$-band images of the five 
fields to allow accurate weak lensing 
studies. It is these $R$-band images,
along with the photometric redshift tables, that we make use 
of in this analysis. 
Gray et al (2002) discuss
the procedure used to reduce the $R$ band imaging data, which totalled
6.3 hours.
The {\tt imcat} shear analysis package was
applied to our reduced image (see Gray et al 2002 for details). This
resulted in a catalogue of galaxies with centroids and shear estimates
throughout our field, corrected for the effects of PSF circularization
and anisotropic smearing.  We appended to this catalogue the
photometric redshifts estimated for each galaxy from the standard
COMBO-17 analysis of the full multicolour dataset.  Wolf et al (2002)
describe in detail the photometric redshift estimation methods used to
obtain typical accuracies of $\sigma_z=0.05$ for galaxies throughout
$0<z<1$.

\subsection{The COMBO17 shear power spectra}

\begin{figure}[t]
\centering
\begin{picture}(200,230)
\includegraphics{comparison.ps}

\includegraphics{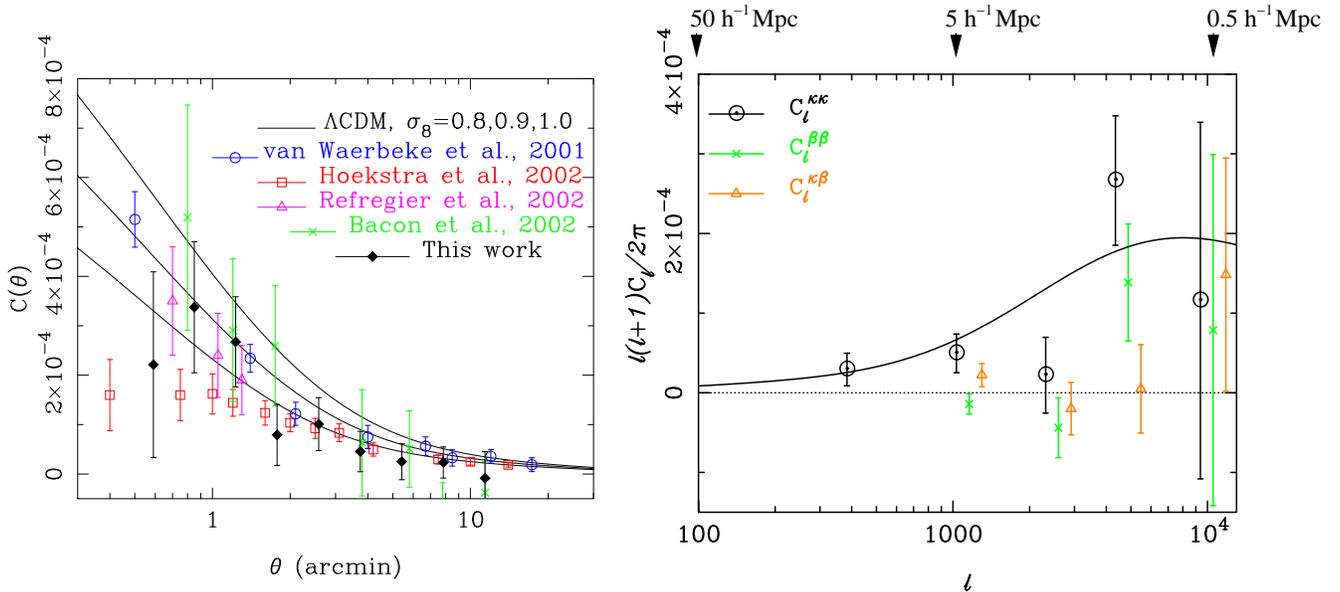}
\end{picture}
\caption{\em {\bf LHS:} The total correlation function, $C(\theta)$, as 
measured from COMBO-17, along with 
the most recent cosmic shear measurements from the four other groups 
indicated. Beyond a scale 
of 1 arcmin, the measurements are in broad agreement. The correlation 
function predictions for
a flat $\Lambda$CDM cosmology, for three values of the power spectrum 
normalisation 
(from top to bottom: $\sigma_8=1.0,0.9,0.8$) are also plotted. 
{\bf RHS:} The cosmic shear power spectra from COMBO-17. 
Plotted on a linear-log scale are 
$C^{\kappa\kappa}_\ell$ (circles), $C^{\beta \beta}_\ell$ (crosses) 
and $C^{\kappa\beta}_\ell$ (triangles) in 5 band-averaged band powers, 
as a function of multipole, $\ell$, estimated from the optimal combination of a
maximum likelihood analysis of the four COMBO-17 fields, CDFS, SGP,
FDF, and S11. The error bars are estimated from the Fisher
matrix. The solid curve is the shear power spectrum expected for
a $\sigma_8=0.8$ normalised $\Lambda$CDM model. Once again, the 
$C^{\beta \beta}_\ell$ and $C^{\kappa \beta}_\ell$ points have been
slightly horizontally displaced for clarity (from Brown et al 2003)} 
\label{comparison}
\end{figure}

To compare our data with other surveys we have estimated the 
shear correlation function,
$C(\theta)$, from COMBO17 and compared these with the results of other
surveys. The results are shown in Figure  \ref{comparison} (LHS),
 where we see broad agreement with other surveys.

Fig. \ref{comparison} (RHS) shows the COMBO17 
shear power from a Maximum Likelihood analysis of the fields. Details 
of the analysis are given in  Brown et al (2003). 
Three of our five band power measurements agree within their error 
bars with the $\sigma_8=0.8$ normalised $\Lambda$CDM model plotted. 
The high level of power measured at $\ell \sim 4000$ seems to be 
present in each of the fields we have analysed. The largest 
discrepancy between the model and the measurements occurs at 
$\ell=2000$ where less than half the power is found. 
We can check our results for systematic effects by estimating the
power in the $\beta$-$\beta$ modes, as well as the
$\kappa$-$\beta$ cross correlation. In all of our band powers the
$\beta$-$\beta$ correlation is below the detected signal in $\kappa$ 
and is consistent with zero in all but one band power at $\ell 
\sim 4000$. Similarly the $\kappa$-$\beta$ cross correlation is well
below our measurement of shear power, and is consistent with zero 
except at $\ell \sim 1000$ where a significant detection appears. 
We conclude from the minimal power found in these spectra that our 
results are not strongly contaminated by systematic effects.
Subsequently we have used the photometric redshift information to remove
close galaxy pairs to remove any intrinsic alignment effects (Heyman et al 
 2003). This had a negligible effect on our results, indicating that 
for current shear surveys this is not a major source of systematic.

\subsubsection{Estimating $\Omega_m$ and $\sigma_8$}
\begin{figure}[t]
\centering
\begin{picture}(190,220)
\includegraphics{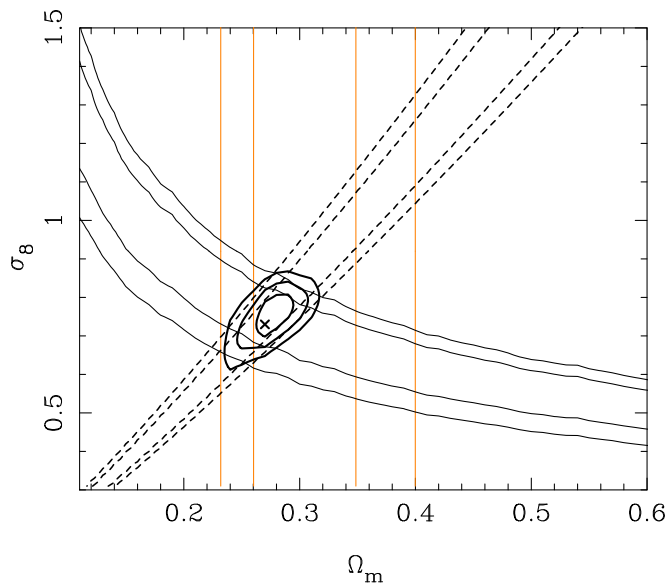}
\end{picture}
\vspace{5mm}
\caption{\em The likelihood surface of $\sigma_8$ and $\Omega_m$ from
combining the COMBO-17 dataset with the 2dFGRS and pre-WMAP CMB constraints. 
The dark thin solid contours are the constraints obtained from our shear power spectrum
analysis described in the previous section. The lighter vertical contours are the 
constraints on $\Omega_m$ obtained from the 2dFGRS where we have applied the priors 
described in the text to the 2dF data. The dashed set of contours
are the constraints from a compilation of six pre-WMAP CMB experiments 
where we have assumed an optical depth to reionization of $\tau=0.10$. 
The dark heavy contours are the 1,2 and 3$\sigma$ combined constraints from the 
three methods. The best fit values of $\Omega_m$ and $\sigma_8$ are also indicated (from Brown et al 2003).} 
\label{params_2dfc17}
\end{figure}

We are now in a 
position to use our measured shear power spectrum estimates 
to obtain a joint measurement of the normalisation of
the mass power spectrum $\sigma_8$, and the matter density
$\Omega_m$, by fitting theoretical shear
correlation functions and power spectra, calculated for particular 
values of these parameters, to our measurements. Details are given 
 in Brown et al (2003).
 Figure \ref{params_2dfc17}
shows the results of our parameter analysis in the $\sigma_8-\Omega_m$
 plane (dark thin solid lines) for COMB017. A best-fit gives 
$\sigma_8 = (0.72\pm 0.09) (\Omega_m/0.3)^{-0.49}$.

\subsubsection{Combination with the 2dFGRS and Pre-WMAP CMB experiments}

We can combine this confidence region with parameter estimations 
from other sources such as
the 2dF Galaxy Redshift Survey (2dFGRS, Percival et al. 2002) and the various 
pre-WMAP CMB experiments (Lewis \& Bridle 2002 \& references therein).
See Brown et al (2003) again for details of this combination.
The results of this combination are shown in Figure \ref{params_2dfc17} for an
optical depth of $\tau=0.10$. Note that the 2dFGRS data we have used for this 
estimation constrains $\Omega_m$ only and so the constraints shown on $\sigma_8$ 
come wholly from the cosmic shear and CMB measurements. We measure 
best-fit values of $\Omega_m=0.27^{+0.02}_{-0.01}$ and  
$\sigma_8=0.73^{+0.06}_{-0.03}$ from the combined data. This compares
well with the WMAP (+CBI, ACBAR, 2dFGRS and Lyman-$\alpha$ surveys) 
results of $\Omega_m=0.27\pm0.04$ and  $\sigma_8=0.84\pm0.04$. Although
 there is some overlap in these surveys there does seem to be good
 convergence in these results. However to push to the even high accuracies 
 we can expect in future lensing surveys, the main sources of systematics 
must be identified and removed.

\section{Mapping the 3-D dark matter}

While the cosmic shear signal, with an accurate 
knowledge of the source redshift distribution, provides an
accurate probe of the projected dark matter component and
cosmological parameters, the combination of shear and source redshifts 
contains much more information. Wittman et al (2001, 2002) have
demonstrated the utility of using source redshifts by
inferring a cluster redshift from shear and photometric redshift
information for galaxies in their sample. The importance of redshift 
information to remove intrinsic galaxy alignments from shear studies has also 
 been discussed by Heymans \& Heavens (2002) and King \& Schneider (2002a,b).
Beyond this one would still 
 like to image the full 3-D dark matter distribution over
 a large fraction of the Hubble volume. In particular 
 this would aid seeing the cosmological growth of structure over
cosmic time -- something that can be calculated in the linear regime
 to high accuracy. In addition, while surveys such as WMAP have helped
pin down cosmological parameter at high redshift, one would like to 
see them evolve. This is particularly important for parameters such as
 the equation of state of the universe, where its evolution can be used
to distinguish between models for the dark energy. In fact it 
 turns out that the 3-D mapping of dark matter is entirely feasible 
with good data. Here we outline the method proposed by Taylor (2001) and 
then an application to the COMBO17 data (see Taylor et al 2003). 
See also Hu \& Keaton (2002)
for a pixelised implementation to simulated data.

\subsection{3-D reconstruction}

\begin{figure}[t]
\centering
\begin{picture}(200,180)
\includegraphics{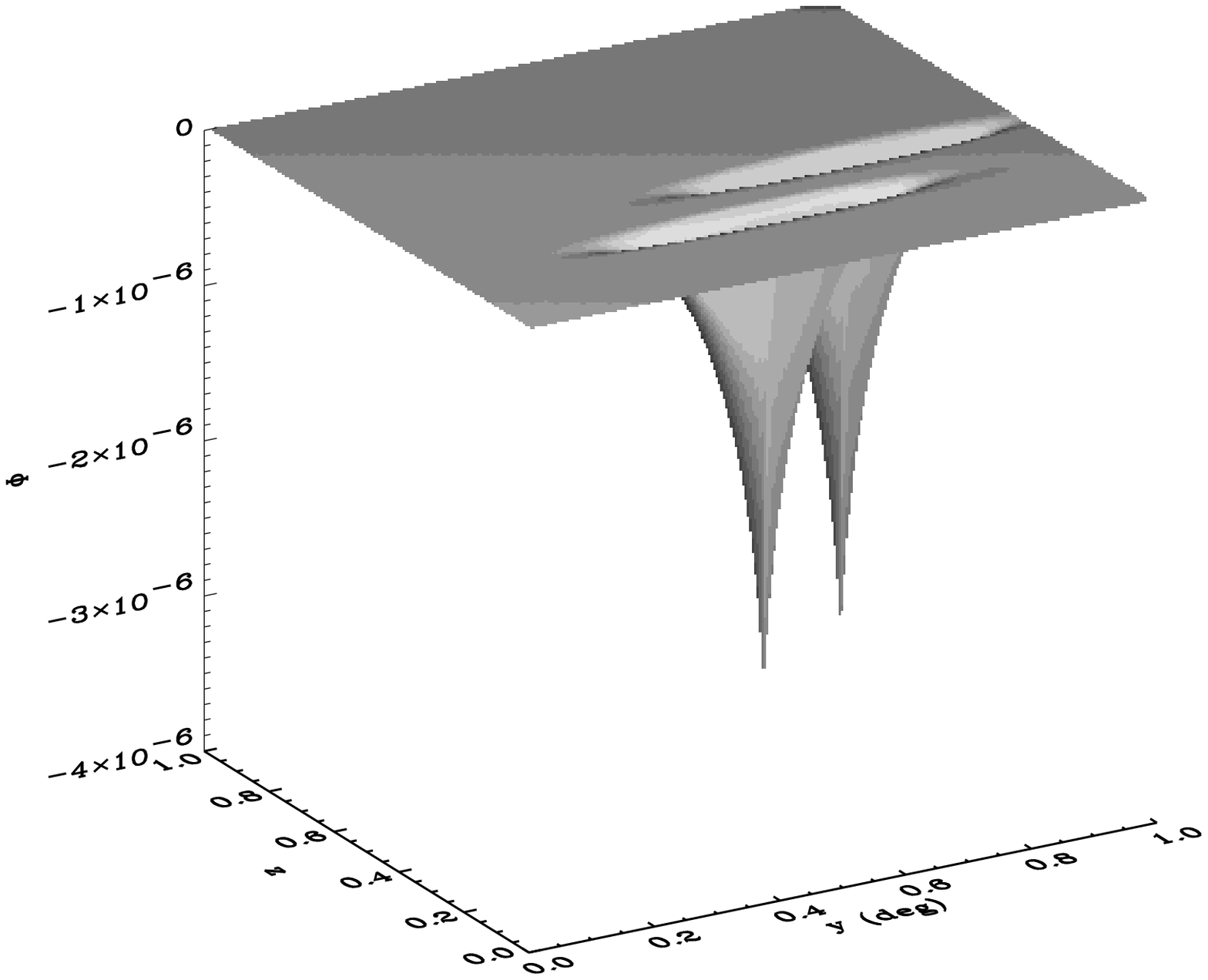}
\includegraphics{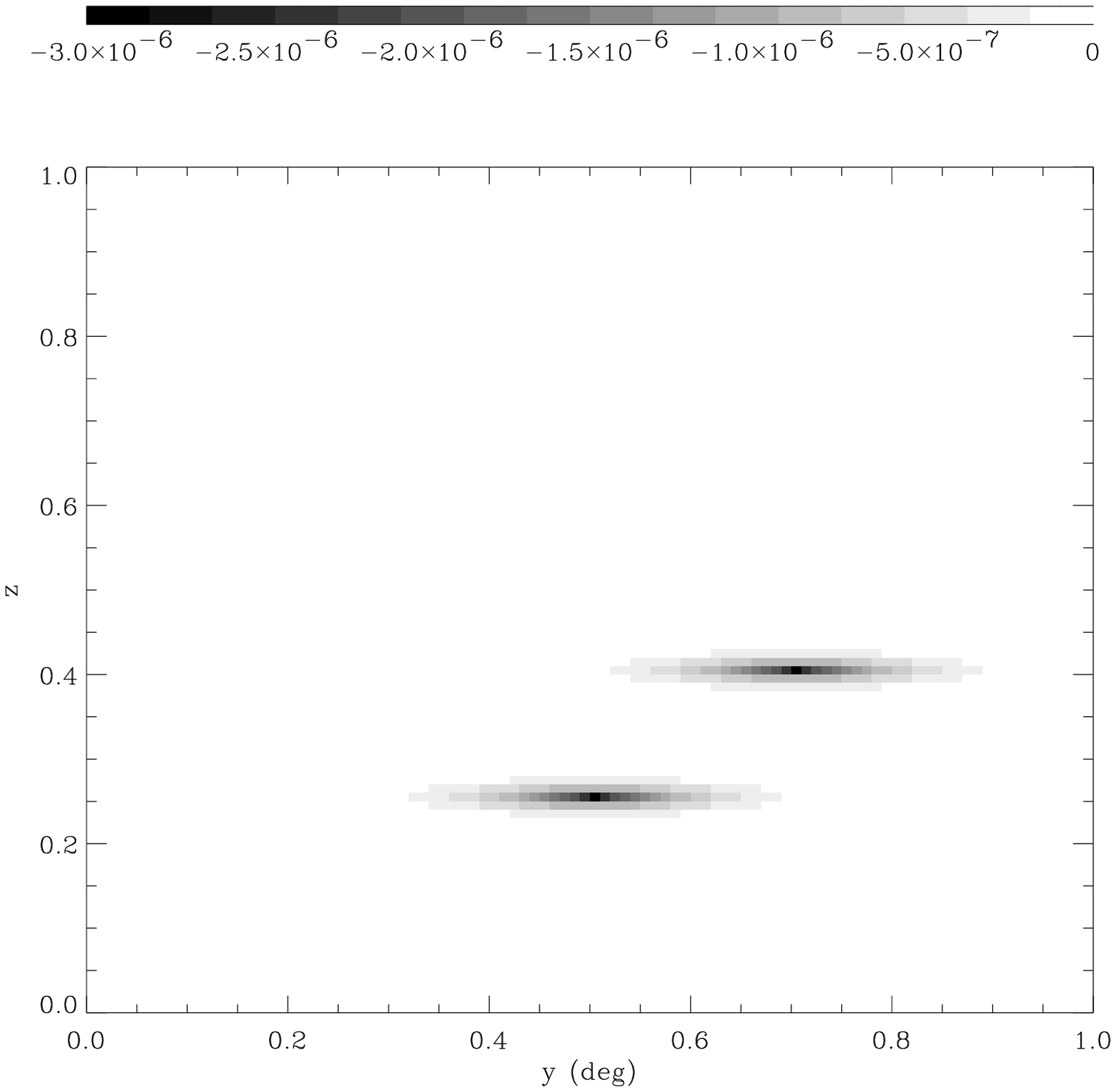}
\end{picture}
\vspace{5mm}
\caption{\em Example of
gravitational potential. Here we simulate the gravitational
potential for two NFW clusters at ${\mathbf r}=(0.5,0.5,0.25)$ and
$(0.5,0.7,0.4)$; the upper panel is a 3-D representation of
($y,z,\Phi$) resulting from an $x$ slice at $x=0.5^\circ$; the
lower panel displays the $\Phi$ values for ($y,z$) in greyscale.}
\label{fig:grav1}
\end{figure}

The inverse relation to equation (\ref{pot}) is
 \be
    \Phi(r,r\thetab) = \frac{1}{2} \de_r r^2 \de_r \,
    \phi(r,r\thetab),
 \label{inv}
 \ee
 which can be verified by substitution into
 equation (\ref{pot}) and integrating by parts. Here the Newtonian
 potential is evaluated at the source position, and
 I assume the lensing potential is smooth enough to allow
 differentiation by the radial derivative $\de_r =
 \thetab.\nabla$.
The shear field provides an estimate of the lensing potential up
to an arbitrary quadratic function;
 \be
    \widehat{\phi}(r,r\thetab) = \phi(r,r\thetab) + \chi(r,\thetab),
  \ee
  where
  \be
    \chi(r,\thetab) = \psi(r) +  \eta(r) \theta_x+\mu(r)\theta_y
    + \nu(r)(\theta_x^2 + \theta_y^2) ,
 \ee
  is a solution to
 \be
    \left(\de_i \de_j - \frac{1}{2} \delta^K_{ij}
    \de^2\right) \chi(r,\thetab) =0 ,
 \ee
 and $\psi$, $\eta$, $\mu$ and $\nu$ are arbitrary radial functions.
 This degeneracy is related to the so-called sheet-mass degeneracy which
 appears as a constant of integration, $4 \nu$, when estimating the
 convergence from the shear field (Falco, Gorenstein \& Shapiro 1985).
 However note that the
 degeneracy we find here will still appear, with $\nu=0$, even if the
 convergence field is used to estimate the lensing potential.
 These terms can be removed by taking moments of the measured lens
 potential over the area of a survey. Defining
 \be
    \chi_{p,q}(r) = \frac{1}{A} \int_A\! d^2\theta \, \widehat{\phi}(r,r\thetab)
    (\theta_x^p + \theta_y^q),
 \ee
 where $A$ is the area of a survey, an estimate of the lensing
 potential, with the mean, gradient and paraboloid contributions removed, is
 \be
    \phi = \widehat{\phi} - \widehat{\chi}
 \ee
 where for a circular survey with radius $\Theta$,
 \be
    \widehat{\chi}=
      4 \chi_{0,0} \left(1-\frac{3 \theta^2}{2 \Theta^2}\right) -
       \frac{6 \chi_{2,2}}{\Theta^2} \left( 1 - \frac{2 \theta^2}{\Theta^2}
       \right)
         + \frac{2}{\Theta^2} \left(\chi_{1,0}\theta_x
    + \chi_{0,1}\theta_y\right),
 \ee
 assuming that the true potential averages to zero.
 Hence
 \be
    \Phi(r,r\thetab) =  \frac{1}{2} \de_r r^2
     \de_r \left(\widehat{\phi}(r,r\thetab) - \widehat{\chi}(r,\thetab)\right)
    \label{unbiaspot}
 \ee
 is an estimate of the Newtonian gravitational potential,
 while
 \be
    \delta(r,r\thetab) = \frac{ \lambda_H^2 a}{3 \Omega_m} \nabla^2 \de_r r^2
     \de_r \left(\widehat{\phi}(r,r\thetab)) - \widehat{\chi}(r,\thetab)\right)
     \label{unbiasdel}
 \ee
 is an estimate of the density field.
 The derivation of equations (\ref{unbiaspot}) and (\ref{unbiasdel})
 are the main results
 of this paper, and demonstrate that the full, 3-D
 Newtonian potential and density fields can be
 reconstructed from weak lensing observations.
 In practice $\chi$ will not be a major problem for
 large surveys where the moments of the true lensing potential
 will average to zero. However for small fields this may not
 be so true, and one may instead wish to set the edge of the
 field to $\phi=0$.

\subsection{Simulations}

Having set out the formal method for a 3-D reconstruction of the
 dark matter distribution, we now test it on simulations. 
 Our simulation consists of a double cluster with a Navarro-Frenk-White
 profile (Figure \ref{fig:grav1}). Further details can be found in Bacon and 
Taylor (2003).

\subsection{Perfect dark matter reconstruction}

\begin{figure}[t]
\centering
\begin{picture}(200,180)

\includegraphics{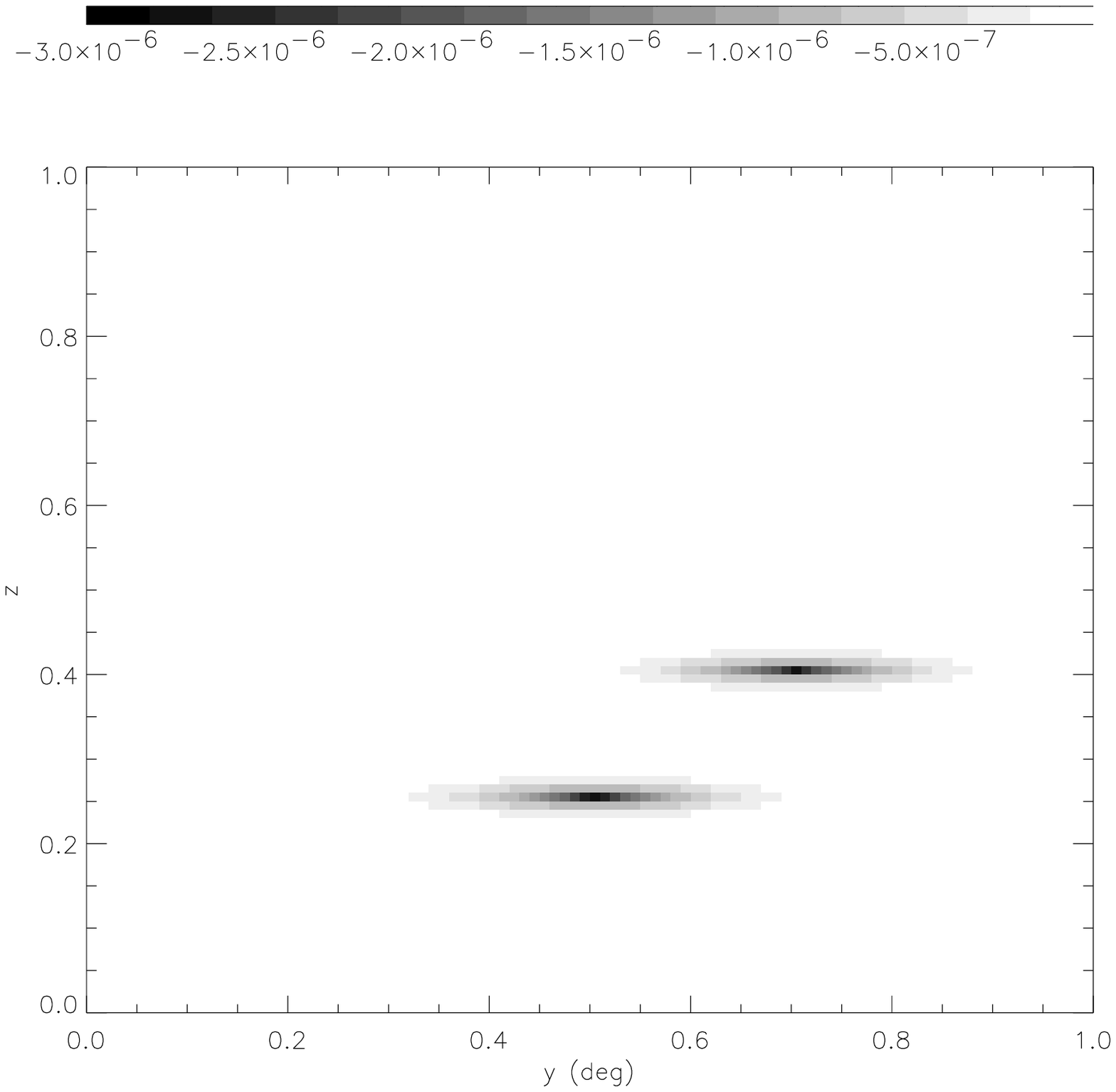}

\includegraphics{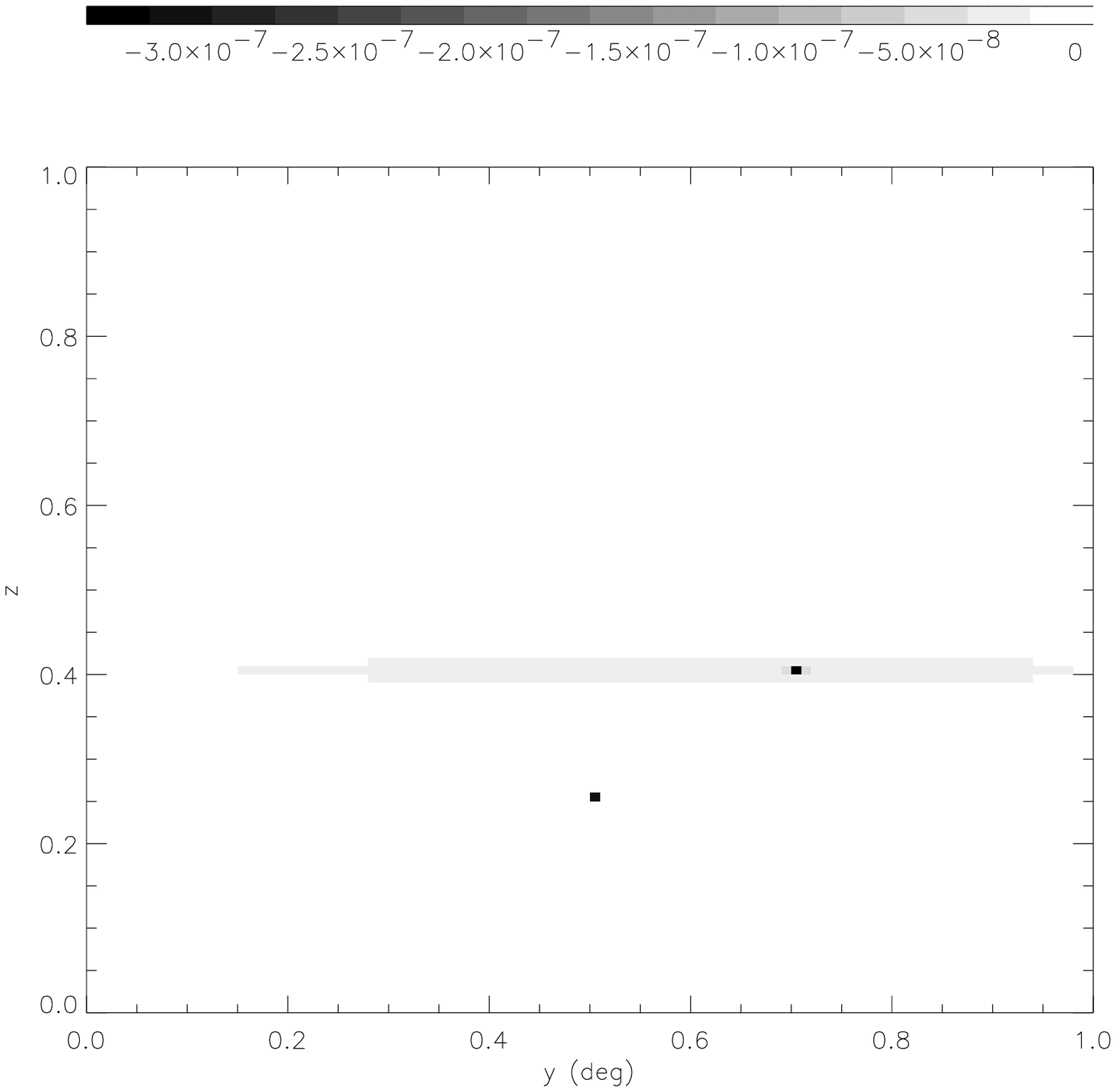}
\end{picture}
\vspace{5mm}
\caption{\em Top panel: reconstructed gravitational potential using the
full shear field of Figure 6, in a $100^3$ grid. This is a
$(y,z)$ plane at $x=0.5^\circ$. Bottom panel: difference between input and
recovered gravitational potential fields.}
\label{fig:grav2}
\end{figure}

First we examine the accuracy of our method when the shear field
is perfectly known everywhere. Figure \ref{fig:grav2} 
displays the reconstructed $\Phi$ and the difference between input and
reconstructed fields. It can be seen that, with perfect
knowledge of the shear field, a good reconstruction is achieved with
our method. For the cluster at $z=0.25$, the error is $<0.4$\% of the
signal within a radius of 0.2 degrees of the cluster, except for the
core pixel where the error is 11.5\%. For the cluster at $z=0.5$, the
error is $<5\%$ of the signal within a radius of 0.1 degrees of the
cluster, except for the core pixel where the error is 17.6\%. This is
again due to the cusp at the cluster centre, which cannot be followed
well by our averaged shear field. Nevertheless, it is clear that our
method is successful in reconstructing cluster gravitational potentials
in the absence of noise.

\subsection{Realistic Dark Matter Reconstructions}

Having demonstrated that the inversions of the shear field to obtain
the gravitational potential is viable in the absence of
noise, we now wish to add the two primary sources of noise present in
lensing experiments: Poisson noise due to only sampling the field at a
finite set of galaxy positions, and additional noise due to galaxies'
non-zero ellipticities. As an example we use number densities for
plausible space-based
experiments (100 per sq arcmin), and 
an appropriate number of objects at each redshift slice.

\begin{figure}
\psfig{figure=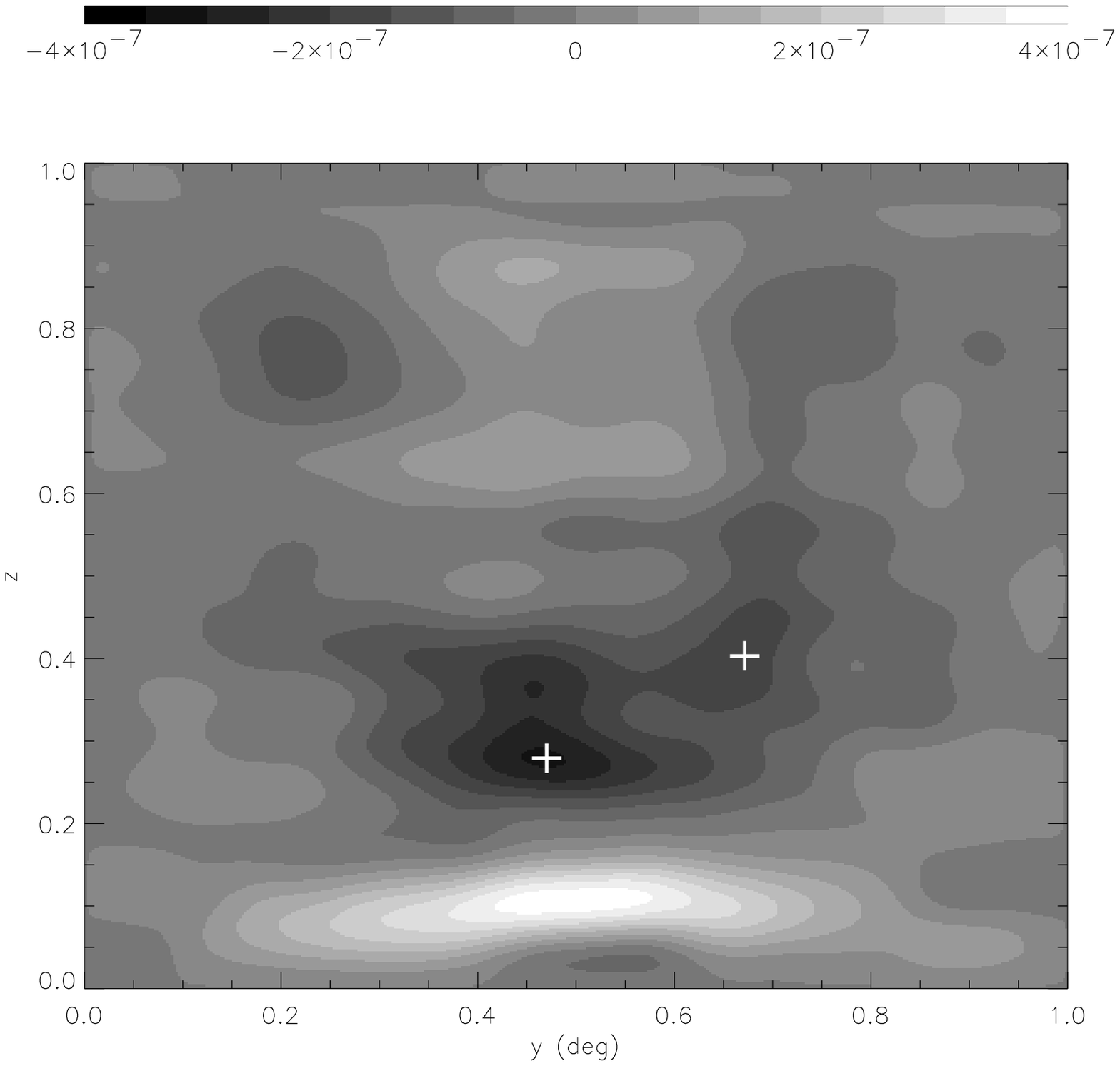,width=80mm}
\includegraphics{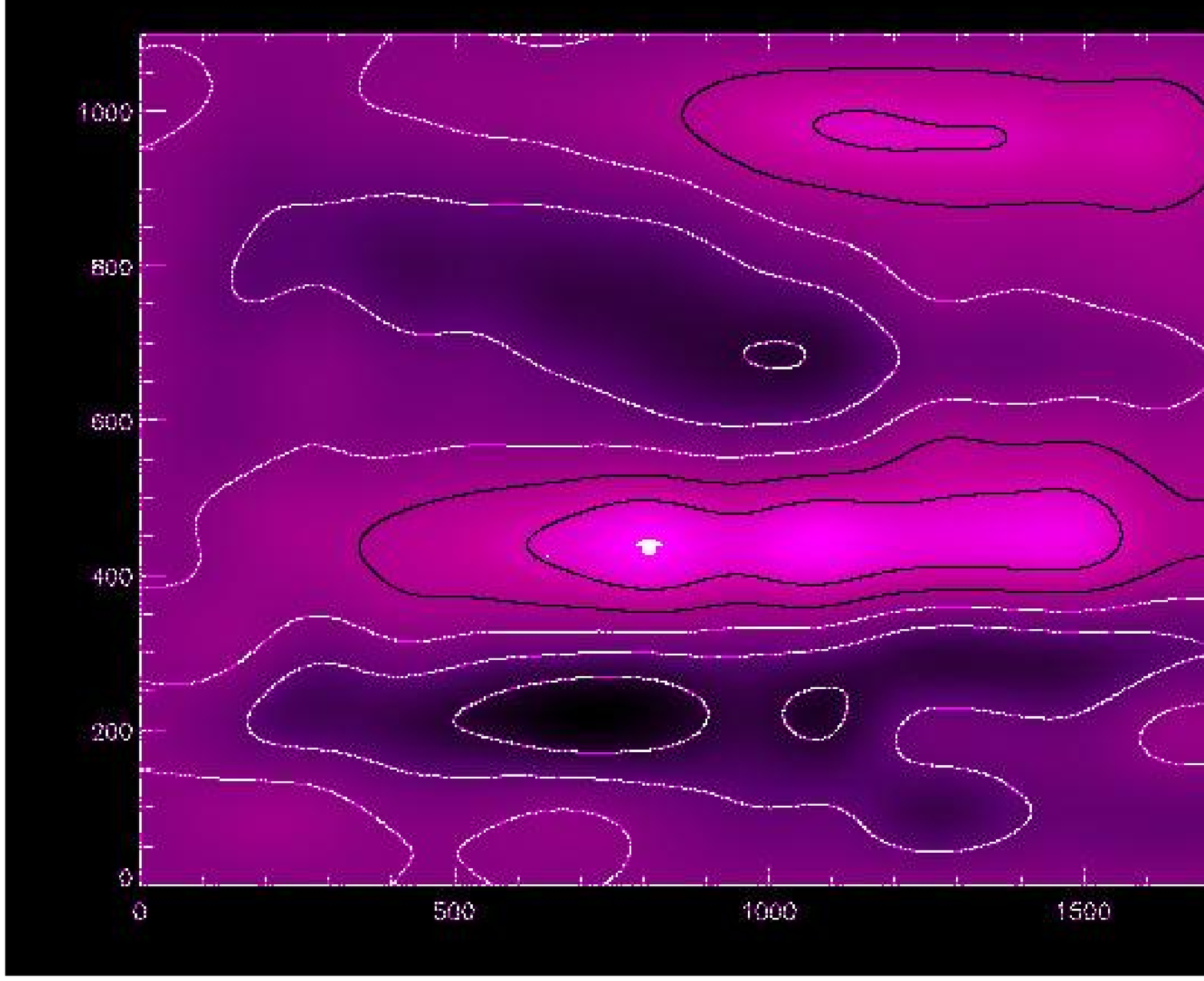}
\caption{\em {\bf LHS} Reconstructed dark matter potential for our space-based
experiment, using galaxy properties as in Figure 9, after Wiener
filtering. Note the detections of clusters at $(y,z)=(0.5,0.2)$ and
$(0.7,0.4)$ with $\nu=4.2$ and 2.1 respectively (crosses show trough
minima).
{\bf RHS}
Cross-section in redshift and angles of the 3-D dark matter potential 
for the COMBO17 A901/2 supercluster field. Note that the lower large 
potential wells, at $z=0.2$, correspond to the A901/2 supercluster, and
the higher potential well, 
at $z=0.4$, corresponds to an unknown background cluster. 
}
\label{simspace}
\end{figure}

Figure \ref{simspace} (LHS) shows the dark matter potential measurements for our
notional space-based survey after Wiener filtering. 
We measure the $z=0.25$ cluster with $\nu=4.2$ at the peak
(trough) pixel of its gravitational potential well, with $\nu=2.1$ at
the gravitational potential trough of the $z=0.4$ cluster. We also
find a substantial noise peak in the foreground at $z=0.1$. While the
recovery of the cluster gravitational potentials therefore constitutes
a challenging measurement, we are indeed able to reconstruct useful
information in the gravitational potential field itself. (The {\em
detection} significance of these clusters is much higher than the
measurement $\nu$ at a given point in the cluster).

\subsection{Reconstructing the Dark Matter Potential in COMBO17}

 In this section, we will
use these methods to calculate the dark matter potential for the
COMBO17 supercluster A901/2 (See Gray et al 2002 for a 2-D shear 
analysis). For more details of the 3-D
dark matter reconstruction see Taylor et al (2003).

Fig.~\ref{simspace} (RHS) shows a cross-section through the final
$\Phi$ field. Note several significant features of this
field. Firstly, we see that the gravitational
potential trough associated with supercluster at $z=0.16$ is clearly
recovered at the bottom of the plot, with a peak pixel S/N of 2.9.
Secondly, further analysis shows that the gravitational
potential displays troughs at approximately the positions of the three
component clusters of A901/2.
Thirdly, we note that a second mass concentration is measured at
$z\simeq0.4$, higher up in the plot. 
This has a peak pixel S/N of 2.8, so is certainly
significant.

While the signal-to-noise on this ground-based application of the
3-D reconstruction is low, this clearly demonstrates the method. 
With higher numbers of sources in space-based surveys, we can expect 
to map out the dark matter distribution in 3-D over a significant 
fraction of the Hubble volume. Such 3-D dark matter catalogues can then be
 used as the basis for constructing mass-selected cluster catalogues
with no projection effects.

\section{Dark Energy from Lens Tomography}

While the combination of shear and redshift information can be use to
directly map the 3-D dark matter, an orthogonal application can 
also be made to probe the dark energy component of the Universe.
The dark energy makes itself felt via its effect on the evolution
of the universe and in particular the Hubble parameter, $H(a)$.
This effects both the evolution of dark matter, and the geometry
of the Universe. This effect has been utilised by, for example, 
Hu (2002), Huterer (2002),
Abazajian \& Dodelson (2002), Heavens (2003), Refregier et al (2003),
Knox (2003) and Linder \& Jenkins (2003). 

Recently Jain \& Taylor (2003) have proposed a new method based purely
on the geometric effect of dark energy on gravitational lensing. 
This involves a particularly simple cross-correlation statistic: 
the average tangential shear around massive foreground halos 
associated with galaxy groups and galaxy clusters. 
We show that cross-correlation tomography measures
ratios of angular diameter distances over a range of redshifts. 
The distances are given by integrals of the expansion rate, which in
turn depends on the equation of state of the dark energy. Thus the
lensing tomography we propose can constrain the evolution of dark energy. 
Lens tomography has also been 
studied as a valuable means of introducing redshift information 
into shear power spectra as a general method for improving parameter
estimation (see e.g. Seljak 1998, Hu 1999, 2002, Huterer 2002, King \& Schneider 
2002b).

\subsection{Method}

The dark energy has equation of
state $p=w\rho$, with $w=-1$ corresponding to a cosmological constant. 
The Hubble parameter $H(a)$ is given by 
\begin{equation}
H(a) = H_0\left[\Omega_{m} a^{-3} + 
\Omega_{\rm de} e^{-3  \int_1^a d \ln a' (1+w(a'))}\right]^{1\over 2} ,
\label{hubble}
\end{equation}
where $H_0$ is the Hubble parameter today. The comoving distance is
$
r(a) = \int_0^a d a'/a'^2 H(a') $.

We consider the lensing induced cross-correlation between massive foreground
halos, which are traced by galaxies, and the tangential shear with respect 
to the halo center (denoted here $\gamma$): 
$\omega_\times(\theta)\equiv \left<\delta n_{\rm f}(\hth)\gamma(\hth^\prime)\right>$
where $n_{\rm f}(\hth)$ is the number density of foreground galaxies with mean
redshift $\langle z_{\rm f}\rangle$, observed in the direction $\hth$ in the
sky and $\delta n_{\rm f}(\hth) \equiv (n_{\rm f}(\hth)-{\bar{n}_{\rm f}})/{\bar{n}_{\rm f}}$. 
The angle between directions $\hth$ and $\hth^\prime$ is $\theta$. 
The cross-correlation is given by (Moessner \& Jain 1998,  
Guzik \& Seljak 2002): 
\be
\omega_\times (\theta)= 6\pi^2 \om \int_0^{r_H} d r
W_{\rm f}(r) \frac{g (r, r_{b})}{a(r)} 
\int_0^\infty dk\, k\, P_{\rm hm}(r, k)\,
J_\mu\left[k r\theta\right] \; , 
\label{omegagl}
\ee
where  $g(r)$  is the lensing geometry averaged over
 the normalized distribution of background galaxies $W_{\rm b}(\chi)$
\begin{equation}
g(r) = r \int_r^{r_H}\! dr' \,
{r' -r \over r'}W_{\rm b}(r')\ ,
\label{gchi}
\end{equation}
 $P_{\rm hm}(r, k)$ is the halo-mass cross-power spectrum, 
and $W_{\rm f}$ is the foreground halo redshift distribution. 
The Bessel function
$J_\mu$ has subscript $\mu=2$ for the tangential shear
and $\mu=0$ for the convergence. 
The measurement of the mean tangential shear around foreground
galaxies is called galaxy-galaxy lensing. We will consider a generalization
of this to massive halos that span galaxy groups and clusters. 
If the foreground sample has a narrow redshift distribution centered
at $r=r_{\rm f}$, then we
can take $W_{\rm f}$ to be a Dirac-delta function and evaluate the integral
over $r$. All terms except $g(r_{\rm f},r_{\rm b})$ are then 
functions of $r_{\rm f}$, the redshift of the lensing mass. 
The coupling of the foreground and background distributions is contained
solely in $g(r_{\rm f},r_{\rm b})$. Hence 
if we take the ratio of the cross-correlation for two background 
populations with mean redshifts $z_1$ and $z_2$, we get
\begin{equation}
\frac{\omega_{\times,1} (\theta)}{\omega_{\times,2} (\theta)}= 
\frac{g_1(r_{\rm f})}{g_2(r_{\rm f})} \approx 
	\frac{(r_1-r_{\rm f})/r_1}{(r_2-r_{\rm f})/ r_2} ,
\label{ratio}
\end{equation}
where $w_1, g_1$ denote the values of the functions for the background
population with mean redshift $z_1$, and the second approximation
is in
 the limit that the background galaxies also 
have a delta-function distribution.
The above equations show that the change in the cross-correlation with
background redshift does not depend on the galaxy-mass 
power spectrum, nor on $\theta$. We can simply use measurements over
a range of $\theta$ to estimate the distance ratio of equation (\ref{ratio})
for each pair of foreground-background redshifts. The distance ratio in
turn depends on the cosmological parameters $\Omega_{\rm de}$, $w$, and
its evolution $w'$.

\subsection{Signal-to-noise estimate}
A simple way to estimate the signal-to-noise for the cross-correlation
approach is to regard the foreground galaxies as providing a template
for the shear fields of the background galaxies. 
For a perfect template (i.e. for high density of foreground 
galaxies and no biasing), the errors are solely due to the finite
intrinsic ellipticities of background galaxies. Thus
the fractional error in our measurement of the background shears is simply
\begin{equation}
\frac{\delta \gamma}{\gamma} \sim 
\frac{\sigma_\epsilon}
{\sqrt{N_{\rm total}} \langle\gamma\rangle_{\rm rms}} 
\sim 0.2 \times 10^{-3} \left(\frac{100}{n_g}
\right)^{1/2} \left(\frac{0.1}{f_{\rm sky}}\right)^{1/2} ,
\label{sn}
\end{equation}
where
the total number of background galaxies is $N_{\rm total} = n_g A = n_g 
f_{\rm sky} A_{\rm sky}$, $A$ is the survey area, the lensing 
induced rms shear $\langle\gamma\rangle_{\rm rms} \simeq 0.04$, 
 the intrinsic ellipticity dispersion $\sigma_\epsilon=0.3$, and
where the number density $n_g$ has units per square arcminute. 
Thus for the fiducial parameters $\fsky=0.1$ and $n_g=100$, one can 
expect to measure the background shear to 0.1\% accuracy at about 
5-$\sigma$. We find that such a signal corresponds to changes in $w$ of 
a few percent; 
hence this is the approximate sensitivity we expect in the absence of 
systematic errors. 

\subsection{Dark energy parameters}

\begin{figure}[t]
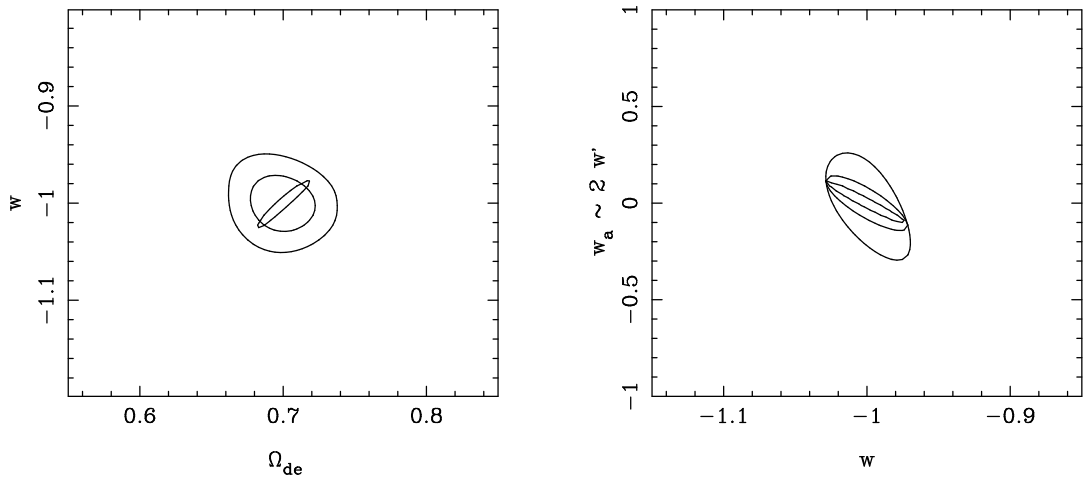

\vspace{7.5cm}
\caption{\em Contours in the $\Omega_{\rm de}-w$ plane for the fiducial 
lensing survey with $\fsky=0.1$. 
The inner contour assumes no evolution of dark energy, $w_a=0$, 
while the two outer contours marginalize over $w_a$, with
external constraints on $\ol$ corresponding to $\sigma(\ol)=0.01, 0.03$ 
(see text). The $68\%$ confidence interval is shown in each of the contours. 
Contours in the $w-w'$ plane for the fiducial 
lensing survey with $\fsky=0.1$, as in Figure \ref{fig:omegaw}.  
The inner contour assumes $\ol=0.7$, while the outer two contours
marginalize over $\ol$ as in Figure \ref{fig:omegaw}. 
Note that the parameter $w_a = 2\ w'$ at $z=1$ (from Jain \& Taylor 2003).
}
\includegraphics{fig1.ps}
\includegraphics{fig2.ps}
\label{fig:omegaw}
\end{figure}

We perform a $\chi^2$ minimization over the mean shear amplitudes at the
two background distributions for each foreground slice to fit for the dark 
energy parameters. The time dependence of $w$ is parameterized as 
$w=w_0+w_a (1-a)$ (Linder 2002). For comparison with other work 
we will compare $w_a$ to $w'$ defined by $w=w_0+w'z$. Since the $w'$ 
parameterization is unsuitable for the large redshift range we 
use, any comparison with it can be made only for a choice of redshift. 
For example, at $z=1$, which is well probed by our method and is of interest
in discriminating dark energy models (Linder 2002), 
$w_a = 2 w'$. 
For foreground slices labeled by index $l$ and two background
samples by $1$ and $2$, the $\chi^2$ is given by 
\begin{equation}
\chi^2 = \sum_{l}\left[1 - \frac{R^0(z_l,z_1,z_2)}
{R(z_l,z_1,z_2)}\right]^2         U_{l} , 
\label{chi2}
 \end{equation}
where $R$ is the distance ratio of equation (\ref{ratio}) for given
values of $\ol, w$ and $w_a$, and $R^0$ is the fiducial model with $\ol=0.7, 
w=-1, w_a=0$. The weights $U_{l}$ are given by 
\begin{equation}
U^{-1}_{l} 
= \frac{\sigma_\epsilon^2}{2 n_1 f_l A\ \langle\gamma\rangle^2_{l1}}
+ \frac{\sigma_\epsilon^2}{2 n_2 f_l A\ \langle\gamma\rangle^2_{l2}} ,
\label{weight}
\end{equation}
where $f_l$ is the fraction of the survey area $A$ covered by halo 
apertures in the $l$-th lens slice. The factor of $2$ in the denominator arises
because we are using only one component of the measured ellipticity
whereas $\sigma_\epsilon^2$ denotes the sum of the variances of both
components. 

The results are shown in Figures \ref{fig:omegaw}.
Figure \ref{fig:omegaw} (LHS) shows the constraints in the $\ol-w$ plane. 
The ellipses show the $68\%$ confidence 
region given by $\Delta\chi^2=2.3$. 
The elongated inner contour is for fixed $w_a=0$. The two outer contours
marginalize over $w_a$ assuming external constraints on $\ol$
from the CMB and other probes, corresponding to $\sigma(\ol)=0.01, 0.03$.

Figure \ref{fig:omegaw} (RHS) shows
the constraints in the $w-w_a$ plane if $\ol$ is fixed, or
marginalized over with $\sigma(\ol)=0.01, 0.03$.
The corresponding accuracy on $w$ and $w'$ can be scaled as 
$\fsky^{-1/2}$. For the case with 
$\sigma(\ol)= 0.03$, we obtain
$\sigma(w)\simeq 0.01\fsky^{-1/2}$ and $\sigma(w_a)\simeq 0.06\fsky^{-1/2}$ 
(at $z=1$ this is equivalent to $\sigma(w')\simeq 0.03\fsky^{-1/2}$). 
Note that the value of $\sigma$ on a parameter 
is given by projecting the $\Delta\chi^2=1$
contour on the parameter axis, which is 
smaller than the projection of the inner contour in 
Figure \ref{fig:omegaw} (RHS) with $\Delta\chi^2=2.3$. 
The scaling with $\fsky$ in the parameter errors comes from the number 
of background galaxies (not sample variance), hence for given $\fsky$ 
varying the depth of the survey scales the errors roughly as $n_g^{-1/2}$. 
The results we have shown are for the fiducial redshift $z=0$. 
A different choice of the fiducial redshift changes the relative 
accuracy on $w$ and $w_a$ somewhat, because the degeneracy direction 
in the three parameters changes. A detailed exploration of different models
of $w(a)$ with finer bins in the background redshift distribution would
be of interest.

\section{Summary}

In this paper I have presented three ways of combining weak 
lens shear data with source depth information. Firstly,
one can use shear and redshifts to tie down the positions of 
source galaxies and remove this as a source of uncertainty in 
the measurement of the shear power spectrum. We have demonstrated
this with an application to the COMBO17 dataset (Brown et al 2003).
Secondly, with source redshifts I have demonstrated on simulations
and with the COMBO17data, that the full 3-D dark matter distribution can 
be mapped (Taylor et al 2003). And finally with source redshifts 
the effect of the dark energy component of the universe can be 
 explored via its effect on the geometry of gravitational lensing
by taking ratios of cross-correlations of lens and source galaxies
(Jain \& Taylor 2003).
Clearly, from these applications and many more, 
the availability of  source redshifts has opened up a whole new 
dimension for gravitational lensing.

\section*{Acknowledgements}
 Many thanks to the conference organisers for making it a great 
 meeting. I also thank my collaborators with whom I have worked
 with on these projects, in particular Meghan Gray, David Bacon, 
 Michael Brown, Alan Heavens, Bhuvnesh Jain and Simon Dye. I addition I thank 
 the COMBO17 team, especially Klaus Meisenheimer, and Chris Wolf,
 and finally Peter Schneider, Yannick Mellier, Alex Refregier, 
 John Peacock, Richard Ellis, Sarah Bridle,
 Catherine Heymans,  Martin White, Wayne Hu and Gary Bernstein for 
 many stimulating discussions about weak lensing.

\section*{References}

\bib Abazajian K.N., Dodelson S., 2002, astro-ph/0212216.

\bib Bacon D., Massey R., Refregier A., Ellis R.S., 2002,
submitted to MNRAS, astro-ph 0203134.

\bib Bacon D., Taylor A.N., 2003, MNRAS in press (astro-ph/0212266)

\bib Bartelmann M.,  Schneider P., 2001, Phys. Rep., 340, 291.

\bib Bonnet H., Mellier Y., Fort B., 1994, ApJL, 427, 83.

\bib Broadhurst T., Taylor A.N., Peacock J., 1995, ApJ, 438, 49.

 \bib Brown M. L., Taylor A. N., Hambly N. C., Dye S., 2002, MNRAS, 333, 501

\bib Brown M.L., Taylor A.N., Bacon D., Gray M., Dye S., Meisenheimer K., Wolf C.,
    2003, MNRAS, 341, 100

\bib Catelan P., Kamionkowski M., Blandford R. D., 2001, MNRAS, 323, 713

\bib Crittenden R., Natarajan P., Pen U., Theuns, T., 
  2001, ApJ, 545, 561

 \bib Croft. R. A. C., Metzler C. A., 2001, ApJ, 545, 561 

\bib Falco E.E., Gorenstein M.V. \& Shapiro I.I., 1985, ApJLett, 289, L1

\bib Fort B., Mellier Y., Dantel-Fort M., 1997, A\&A, 321, 353

\bib Gray M. E., Taylor A. N., Meisenheimer K., Dye S., Wolf C., Thommes E., 2002, ApJ, 568, 141.

\bib Guzik J., Seljak U., 2002, MNRAS, 335, 311.

\bib Heavens A. F., 2003, submitted to MNRAS, astro-ph/0304151 

\bib Heavens A. F., Refregier A., Heymans C., 2000, MNRAS, 319, 649

\bib Heymans C., Heavens A. F., 2003, MNRAS, 339, 711 

\bib Heymans C., Brown M., Heavens A.F., Meisenheimer K., Taylor A.N., 
	C. Wolf, 2003, submitted MNRAS

\bib Hoekstra H., Franx M., Kuijken K., Squires G., 1998, ApJ,
504, 636.

\bib Hoekstra H., Yee H., Gladders M., Barrientos L. F.,
Hall P., Infante L., 2002, ApJ, 575, 55.

\bib Hu W., 1999, ApJL, 522, 21.

\bib  Hu W., 2002, Phys.Rev.D,  66, 083515

 \bib Hu W., White M., 2001, ApJ, 554, 67 (HW)

\bib Hu W., Keeton C. R., 2002, submitted to PRD, astro-ph 0205412.

\bib Huterer D., 2002,  PRD, 65, 063001

\bib Jain B., Taylor A.N., 2003, submitted to PRL, astro-ph/0306046

\bib Jarvis M.,  Bernstein G., Jain B., Fischer P., Smith D.,
  Tyson J.A.,  Wittman D., 2003, ApJ, 125, 1014

\bib Kaiser N., Squires G., 1993, ApJ, 404, 441.

\bib King L., Schneider P., 2002a, accepted by A\&A,
astro-ph 0208256.

\bib King L., Schneider P., 2002b, A\&A in press,
astro-ph 0209474.

\bib Knox L., 2003, astro-ph/0304370.

 \bib Lewis A., Bridle S., 2002, Phys. Rev. D., 66, 103511

\bib Linder E.V., arXiv:astro-ph/0210217.

\bib Linder E.V. \& Jenkins A., 2003, astro-ph/0305286.

\bib Luppino G. A., Kaiser N., 1997, ApJ, 475, 20.

\bib Mellier Y., 1999, ARA\&A, 37, 127

\bib Moessner R., Jain B., 1998, MNRAS, 294, L18

 \bib Percival W. J. et al., 2002, MNRAS, 337, 1068     

\bib Refregier A., et al., 2003, astro-ph/0304419.

\bib Squires G., Kaiser N., Fahlman G., Babul A., Woods D., 1996, ApJ, 469, 73.

\bib Seljak U., 1998, ApJ, 506, 64.

\bib Taylor A.N., et al, 1998, ApJ, 501, 539

\bib Taylor A.N., 2001, Phys. Rev. Lett, 
  submitted (astro-ph 0111605)

\bib Taylor A.N., Bacon D.J., Gray M.E., Wolf C., Meisenheimer K., Dye S., 2003,
	submitted MNRAS

\bib Tyson J.A., Valdes F., Wenk R.A., 1990, ApJLett, 349, L1

\bib  van Waerbeke L., Mellier Y., Radovich M., Bertin E.,
Dantel-Fort M., McCracken H. J., Le Fevre O., Foucaud S., Cuillandre
J.-C., Erben T., Jain B., Schneider P., Bernardeau F., Fort B., 2001,
A\&A, 374, 757.
y
\bib Wittman D.M., et al, 2001, ApJ, 557, 89

 \bib Wolf C., Dye S., Kleinheinrich M., Rix H.-W., Meisenheimer K.,
  Wisotzki L., 2001, A\&A, 377, 442   

 \bib Wolf C., Meisenheimer K., Rix H.-W., Borch A., Dye S., 
  Kleinheinrich M., 2003, A\&A, 401, 73   

\end{document}